\documentclass{ws-procs9x6}

\begin{document}

\title{Earth Flyby and Pioneer Anomalies}

\author{M.B. Gerrard\footnote {Corresponding author:- Michael.Gerrard@partnershipsuk.org.uk}  \ and T.J. Sumner}

\address{Department of Physics, Imperial College London, Prince Consort Road,
London. SW7 2BW, UK. E-mail: t.sumner@imperial.ac.uk}


\maketitle

\abstracts{Applying Newtonian dynamics in five dimensions rather
than four, to a universe that is closed, isotropic and expanding,
suggests that under certain circumstances an additional and
previously unidentified acceleration, $a_A$, can arise affecting the four
dimensional motion of spacecraft.  The two cases of this acceleration 
being either real or virtual are considered.  In the real case, simple estimates of $a_A$ are shown to be in partial agreement with reported acceleration
anomalies from several Earth flybys and from the Pioneer spacecraft.   However,
these estimates do not fully reconcile with radio Doppler tracking data.  The virtual case, by contrast, appears to overcome these and other  difficulties with the real case, and is discussed in an addendum.  Furthermore, the virtual case has an altitude dependence which makes detection of any anomaly unlikely above $\sim$2000\,km. Equations governing this additional acceleration have been derived from first principles, without the introduction of free parameters or new constants and without
amendment to the law of gravity.}

\section{Introduction}\label{intro}
Measurements of the velocity profiles of several spacecraft during
Earth flybys, and of Pioneers 10 and 11 following their flybys of,
respectively, Jupiter and Saturn have identified small discrepancies
between expected and observed velocities [1,2]. In particular for
Earth flybys, relative discrepancies in velocity of the order
$10^{-6}$ (which can be either positive or negative) have been
reported and, for the two Pioneer spacecraft, a near constant
acceleration towards the Sun of the order $8.74\times
10^{-10}$\,ms$^{-2}$ has been detected. Currently, these anomalies
are without widely accepted explanations.

The purpose of this paper is to show that the introduction of a
fifth large-scale dimension to the Newtonian equations of motion may
simultaneously account for both these anomalies.

\section{A Five Dimensional Framework}\label{AFDF}
An additional large-scale dimension, $r$, that is orthogonal to the
three space dimensions $s(x,y,z)$ and to the time dimension, $t$, is
introduced and identified as the radius of curvature of
four dimensional space-time in a closed, isotropic and expanding
universe. The dimension $r$ is defined to have units of metres and a
background value remote from gravitating bodies of $r_u$. The
numerical value of $r_u$ is not required for the current analysis,
although one has been proposed by the authors [3]. Radiation freely
propagating at the speed of light, $c$, in this closed isotropic universe will experience an
acceleration, $a_r$, in the direction of $r$ and varying as
$a_r=c^2/r$, for a five dimensional universe that is locally flat
(see Fig. 1)\footnote{Similarities have been identified by the authors between $a_r$ and the MOND parameter $a_o$ [3,4]}. To assist visualisation of the relationship between
space, time and the additional dimension, $r$, only a single space
dimension, $s$, is shown in Fig. 1. Using spherical polar
co-ordinates, the relationships between $s$, $t$ and $r$ are given
by:
\begin{equation}
r_s = r\sin{\alpha}
\end{equation}
\begin{equation}
s = \phi r_s
\end{equation}
\begin{equation}
cdt = rd\alpha
\end{equation}
From these three equations it is possible to derive an expression
for the background expansion of an arc of space, $s$, as follows:
\begin{equation}
\frac{\partial s}{\partial t} = s\frac{c}{r}\cot{\alpha}
\end{equation}

\begin{figure}[t]
\begin{center}
\includegraphics[height=3.5in]{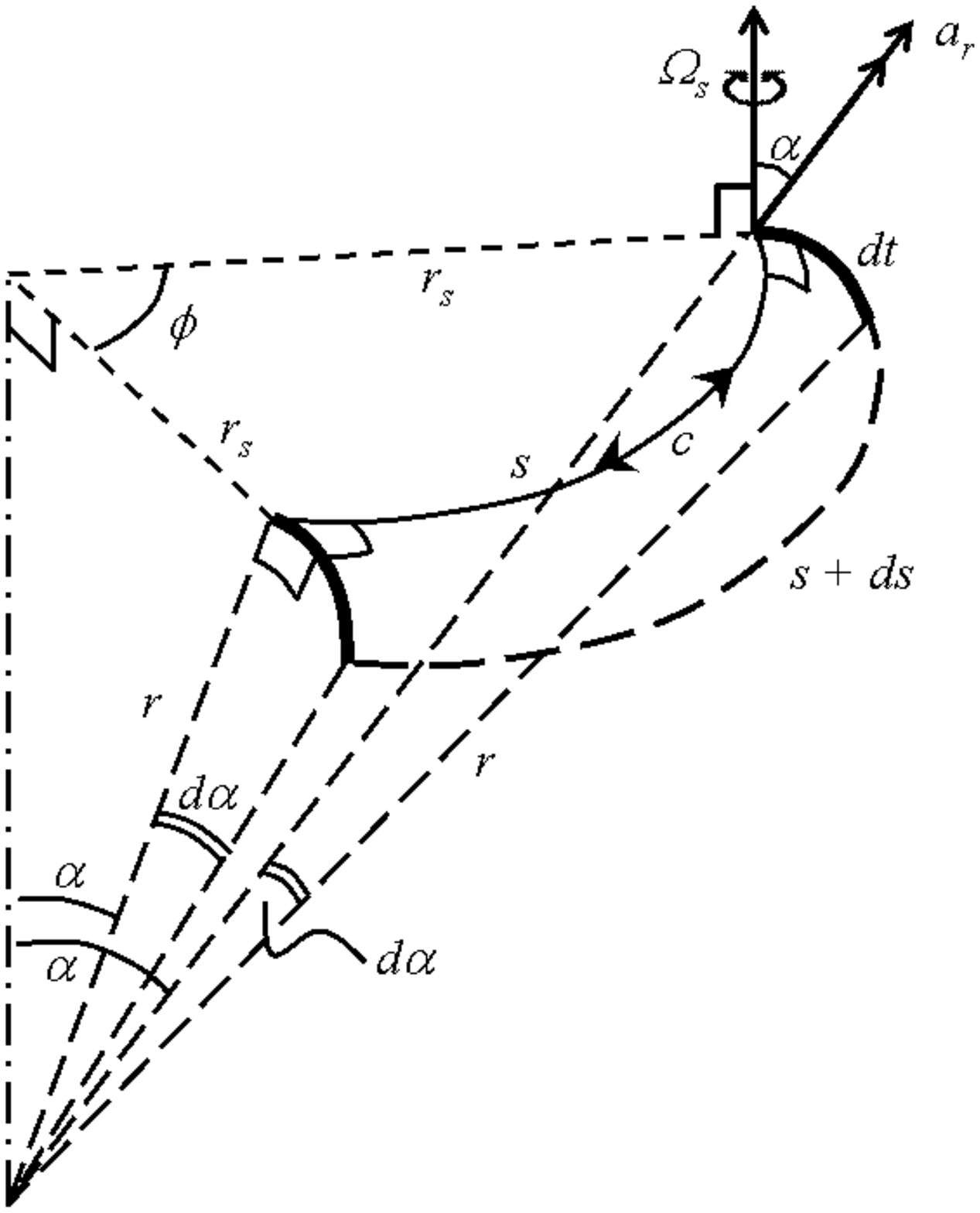}
\vspace{-30mm}
\caption{An arc of space, $s$, in a closed, isotropic and expanding
universe} \label{fig1}
\end{center}
\end{figure}
For each point in space, an angular velocity vector $\Omega_s$ can
be defined which subtends an angle $\alpha$ relative to the
dimension $r$ and whose value is given by equation (5) in which $c$ is defined to be positive:
\begin{equation}
\Omega_s
 = \frac{c}{r_s}
\end{equation}

The presence of a gravitating body of mass $M$ causes the radius of
curvature of space-time, $r$, to reduce relative to its limiting
value $r_u$, according to the following approximation [3]:
\begin{equation}
r \simeq r_u\left(1-\frac{GM}{c^2s}\right)
\end{equation}

The ($s$,$r$) contour defined by this equation is the boundary between four dimensional
space-time and the fifth dimension $r$. It is the locus of points
along which the resolved accelerations of gravity $g_s$ (towards
$M$) and of $a_r$ (in the direction of $r$ increasing) are equal and
opposite.  Hence for $r \simeq r_u$:
\begin{equation}
g_s = -a_r \frac{dr}{ds} = -\frac{GM}{s^2}
\end{equation}

\begin{figure}[h]
\begin{center}
\includegraphics[height=5in,clip=]{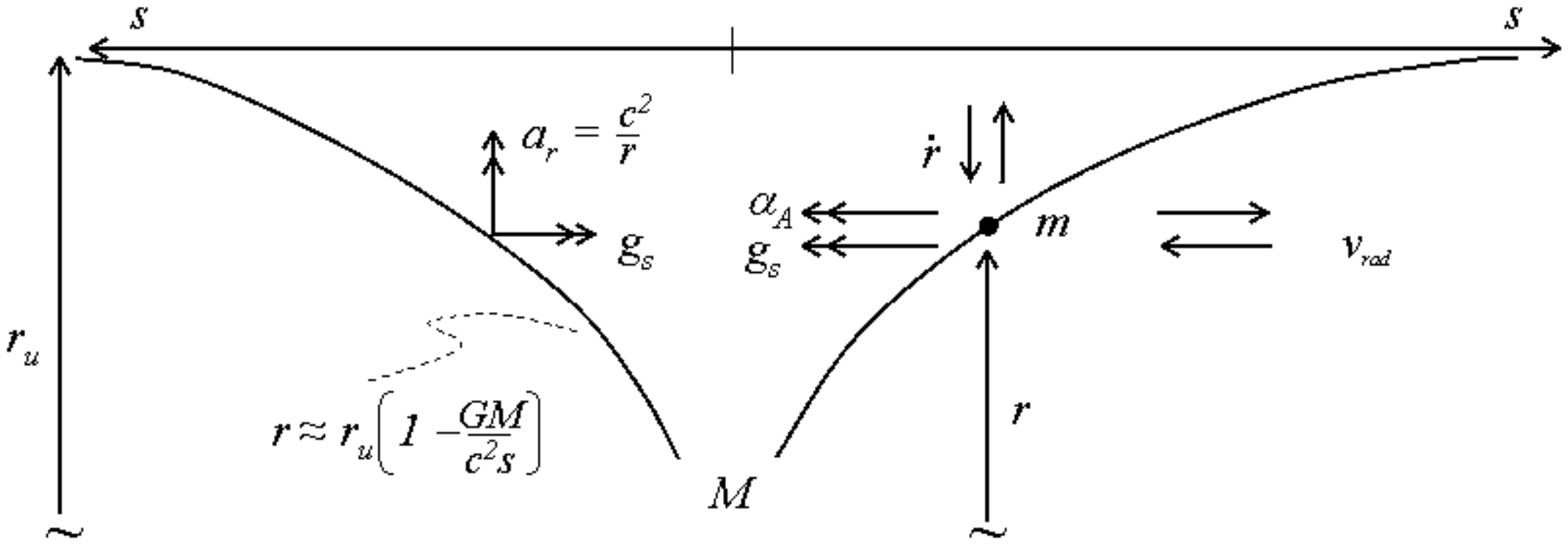}
\vspace*{-90mm}
\caption{The influence of a gravitating body, $M$, on the radius of
curvature of space-time, $r$, and motion of a small body, $m$, along the associated contour $r\left(M,s\right)$} \label{fig2}
\end{center}
\end{figure}

For a small body (shown as $m$ in Fig. 2) travelling at a relative
radial velocity, $v_{rad}$, with respect to $M$, the derivative of
the angular velocity vector $\Omega_s$ is given by equation (8). At
this point in our analysis only variations in $r_s$ due to the
motion of the body $m$ through the gravitational field $g_s$ are
considered (i.e. $\alpha$ is assumed constant, so ignoring the
expansion of the universe):
\begin{equation}
\dot{\Omega}_s = \frac{1}{r_s} g_s \frac{v_{rad}}{c}
\end{equation}

$\dot{\Omega}_s$ gives rise to an additional acceleration $a_A$ in
the space dimension, $s$, as defined by the vector cross product in
equation (9):
\begin{equation}
a_A = r_s \wedge \dot{\Omega}_s
\end{equation}

Substituting from equation (8) provides the following expression for
the acceleration $a_A$:
\begin{equation}
a_A = g_s \frac {v_{rad}}{c}
\end{equation}

$a_A$ acts along the line between $M$ and $m$,  i.e. is co-aligned
with $g_s$.  It may alternatively be expressed, by substituting from equation (7), solely in terms of the fifth dimensional parameter, $r$, as shown in equation (11) from
which it can be seen that $a_A$ is a five dimensional analogue to
the Coriolis acceleration of four dimensional dynamics.
\begin{equation}
a_A = -\frac {c}{r} \wedge \dot{r}
\end{equation}
The additional acceleration, $a_A$, is not a modification to the
gravitational vector $g_s$, so any change in the kinetic energy of
the mass $m$ under the action of $a_A$ is not offset by a reduction
or gain in gravitational potential energy of $m$ with respect to
$M$. Accordingly, the conservation of energy requires that the
action of $a_A$ will only give rise to an observed change in the
kinetic energy of $m$ where energy may be transferred to/from $m$ by
virtue of the general dynamics of the system within which it is
moving, for example in the circumstances of flybys.

\section{Earth Flybys}
It is straightforward to show from equation (10) that where energy
transfer to/from $m$ can take place, the additional change in
velocity due to $a_A$ as $m$ moves radially with respect to $M$
between $s_1$ and $s_2$, is given by equation (12), where
$\overline{\theta}$ is the mean value (between $s_1$ and $s_2$) of
the angle subtended by the radial velocity vector, $v_{rad}$, to the
direction of motion of the spacecraft, such that $\Delta v = \Delta
v_{rad}\cos{\overline{\theta}}$ - noting that the
acceleration $a_A$ and the radial vector are always aligned (see
Fig. 3):

\begin{equation}
\Delta v = \frac{GM}{c} \left(\frac{1}{s_1} - \frac{1}{s_2} \right)
\cos{\overline{\theta}}
\end{equation}

However, in the case of planetary flybys, it is the vector
cross-product between the angular momentum of the planet, in this
case the Earth, around the Sun and the motion of the spacecraft
which defines the energy transfer. Accordingly, the velocity vector
of the spacecraft ($v$) must be resolved into the plane of the
ecliptic and it follows that the cosine of the angle $\beta^\prime$
- the ecliptic angle of incidence of the velocity vector ($v$)  -
determines the extent to which energy transfer takes place as a
consequence of $a_A$ acting on the
spacecraft. Similarly, the angle of divergence
$\overline{\theta^\prime}$ between the velocity ($v$) and radial
velocity ($v_{rad}$) vectors projected into the ecliptic (rather
than simply $\overline{\theta}$) determines the relationship between
 $a_A$ and changes in velocity of the
spacecraft, such that $\Delta v \propto
a_A\cos{\beta^\prime}\cos{\overline{\theta^\prime}}$. The change in
velocity ($\Delta v$) is then given by equation (13) where $M_e$ is
the mass of the Earth:
\begin{equation}
\Delta v = \frac {GM_e}{c}\left(
\frac{1}{s_1}-\frac{1}{s_2}\right)\cos{\beta^\prime}\cos{\overline{\theta^\prime}}
\end{equation}

For an inbound trajectory towards Earth, $\Delta v_i$ will be
positive and, conversely, for an outbound trajectory $\Delta v_o$
will be negative. Hence, the net gain/loss in kinetic energy of the
spacecraft during a flyby due to $a_A$ will be given by the
aggregate of inbound and outbound differences in velocity (i.e.
$\Delta v_i + \Delta v_o$). For a flyby measured over a distance
from/to $s$ via the distance of closest approach (given the symbol
$s_f$), the additional velocity gain/loss on flyby is given by:

\begin{equation}
\Delta v = \Delta v_i + \Delta v_o = \frac {GM_e}{c}\left(
\frac{1}{s_f}-\frac{1}{s}\right)\left(\cos{\beta^\prime_i}\cos{\overline{\theta^\prime}_i}-\cos{\beta^\prime_o}\cos{\overline{\theta^\prime}_o}\right)
\end{equation}

There is an important short-coming to this expression in that during
the flyby the angle of incidence $\beta^\prime$ of the velocity
vector to the ecliptic is not constant, particularly during the transition from
inbound to outbound trajectories. This will tend to introduce
significant inaccuracy for trajectories that are highly asymmetrical
about the plane of the ecliptic. By the same token, use of an
average value, $\overline{\theta^\prime}$, rather than the full
time-dependent profile $\theta^\prime\left(t\right)$ during the
flyby will also be a source of inaccuracy.  The characteristics of
equation (14) are such that for the six reported Earth flybys
(Galileo(1990), Galileo(1992), NEAR(1998), Cassini(1999),
Rosetta(2005) and MESSENGER(2005)), the key portion of the
trajectories which gives rise to the net additional velocity
gain/loss $\Delta v$ is roughly 2 hours either side of the closest
approach.  This arises for two reasons: first, the magnitude of $a_A$ falls away with distance from Earth
($s$) and with a declining radial velocity ($v_{rad}$) of the
spacecraft, such that for each of the inbound and outbound
trajectories, the change in velocity ($\Delta v$) over the 2 hour
segment measured from  the closest approach, is circa 90\% of that
arising over the respective 4 hour interval (and is circa 80\% of
that arising over the respective 24 hour interval); and second,
$a_A$ influences the velocity of the spacecraft ($v$) via the
assymetric time dependent profile of the vector cross-product of the
flyby itself.

\begin{figure}[t]
\begin{center}
\includegraphics[height=4.5in,clip=]{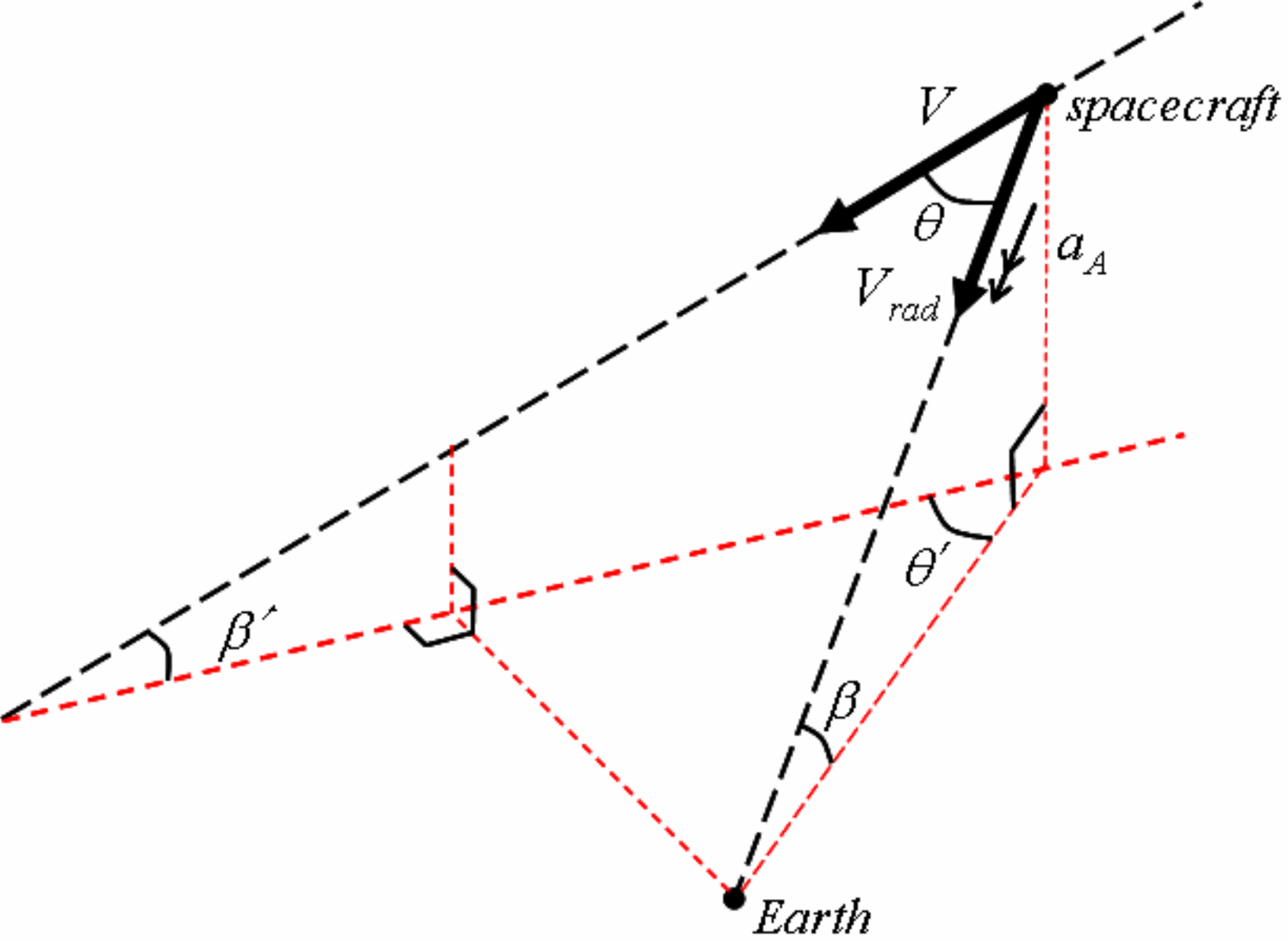}
\vspace*{-65mm}
\caption{Divergence between velocity, $v$, and radial velocity,
$v_{rad}$ vectors during flybys} \label{fig3}
\end{center}
\end{figure}

To take proper account of the time dependency of $\beta^\prime
\left( t\right)$ and $\theta^\prime\left( t\right)$ requires
detailed numerical integration and modelling of the trajectories
during the sectors +/- 2 hours either side of the closest approach,
which is beyond the scope of this paper. However, a simple
approximation can be introduced by splitting the inbound and
outbound calculation into just two sectors each - the first is from
the 2hr boundary location, $s_{2Hr}$, in to $2s_f$ (the outer
sector), and the second for distances $s_f < s < 2s_f$ (the inner
sector). In the outer sector, $\overline{\theta}$ and so
$\overline{\theta^\prime}$ is assumed zero for all inbound and
outbound trajectories and in the inner sector during which $30^o <
\theta^\prime < 90^o$, $\overline{\theta^\prime}$ is set to an
average value of $60^o$ except where a trajectory has a significant
value of $\beta^\prime$. In these cases only (NEAR outbound and
MESSENGER inbound) a reduction in $\overline{\theta^\prime}$ needs
to be made and a value of $30^o$ has been assumed. Also for these
two trajectories (NEAR and MESSENGER) $\beta^\prime$ within the
inner sector is given a value half that of the outer sector, as
appropriate, to allow for the highly asymmetric transition of
$\beta^\prime$ at the closest approach in these cases.

\begin{table}
  \title{\bf{Table I: Comparison of predicted and observed values for
additional velocity gained/lost on flyby}}\label{tab}
\begin{center}
\begin{tabular}{|c|c|c|c|c|c|c|c|}
  \hline
Parameter & Units & Galileo & Galileo & NEAR & Cassini & Rosetta & M'GER \\
\hline \hline Date & & 1990 & 1992 & 1998 & 1999 & 2005
&2005\\\hline $H_{perigee}$ & km & 960 & 303 & 539 & 1175 & 1956 &
2347\\ \hline $s_f$ & km & 7331 & 6674 & 6910 & 7546 & 8327 & 8718
\\ \hline
$s_{2Hr}$ & km$\times 10^{-5}$ & 0.73 & 0.73 & 0.61 & 1.19 & 0.46 & 0.47 \\
\hline $v_f$ & kms$^{-1}$ & 13.740 & 14.080 & 12.739 & 19.026 &
10.517 & 10.389 \\ \hline $\beta^\prime_i$ & Degree & 10.89 & -17.89
& 2.44 & -2.12 & 2.89 & 52.37 \\ \hline $\beta^\prime_o$ & Degree &
-17.66 & -6.71 & -59.86 & -1.62 & -12.43 & -13.73 \\ \hline $\Delta
v$ & mms$^{-1}$ & 3 & -5 & 13 & -0.03 & 2 & [5] \\ \hline
$\Delta v_{obs}$ & mms$^{-1}$ & 3.92 & -4.6 & 13.46 & $<-0.2$ & 1.8 & 0.02 \\
  \hline
\end{tabular}
\end{center}
\end{table}
As can be seen from Table I, \footnote{Values for: (i) the altitude
of closest approach ($H_{perigee}$); (ii) the velocity at closest
approach ($v_f$); and (iii) the measured values for the additional
velocity ($\Delta v_{obs}$) are all taken from [1] with the
exception of the reported value of $\Delta v_{obs}$ for Cassini
which is taken from [2]. Other values in the table are derived as
follows: (a) $s_f$ is the sum of $H_{perigee}$ and the Earth's mean
radius $R_{Earth}$ (6371\,km); (b) the distance of the spacecraft
from Earth approximately 2hr either side of the flyby ($s_{2Hr}$) is taken from [5]; (c) the
ecliptic angle of incidence ($\beta^\prime$) of the velocity vector
is approximated by the ecliptic latitude ($\beta$) of the spacecraft
where the angle of divergence ($\theta$) between the velocity ($v$)
and radial velocity ($v_{rad}$) vectors is close to zero (i.e. other
than during the close flyby itself), where ($\beta$) is calculated
from the equatorial latitude, the right ascension [1] and the
obliquity of the ecliptic ($23.45^o$); (d) $\Delta v$ is calculated
from equation (14) using a two sector approximation as explained in
the main text.} equation (14) applied for simplified two-sector
trajectories predicts values for $\Delta v$ which are in reasonable
agreement (both in regards to magnitude and sign) with most observations
for the spacecraft. The degree of agreement can be further illustrated by solving for a common distance ($s$) which, if applied in equation (14) instead of the common time-cutoff ($\pm2$hrs either side of the perigee) would limit the differences between $\Delta v$ and $\Delta v_{obs}$, particularly for MESSENGER.  This common distance ($s$) lies in the range $72-76\times 10^3$ km and provides agreement between $\Delta v$ and $\Delta v_{obs}$ to an accuracy of $\pm$1 mm/s for all spacecraft\footnote{In fact, the functional form of equation (14) may give rise to better fits of the Doppler tracking data to the equations of 4-dimensional dynamics than expected, notwithstanding the action of $a_A$, for the same reason that equation (22) does in the case of $a_A$ being interpreted as virtual (see section 6.3). }. 

Alternatively, if during the inner sector for all six
trajectories, account is taken of the change in $\beta^\prime$ at
the closest approach as the spacecraft transition from inbound to
outbound trajectories, typically by assuming, where appropriate,
that the average value for $\beta^\prime$ is reduced by 50\% then
close agreement is achieved between observed and predicted values
for $\Delta v$ with values of $\overline{\theta^\prime}$ within
$60^o\pm 2^o$ in the inner sector, except in the case of NEAR and
MESSENGER for which $\overline{\theta^\prime}_o$ and
$\overline{\theta^\prime}_i$ respectively are reduced to $30^o\pm
5^o$ to account for the significant ecliptic angles of incidence of
these trajectories as noted earlier.

\section{Pioneer Anomaly}
Returning to equation (5) to apply this to the Pioneer Anomaly, we
need to generate the time derivative of the angular velocity vector,
$\Omega_s$, including not only the effect of a body moving through a
gravitational field ($g_s$), but also the variation of $r_s$ (and so
of $\Omega_s$) with time, due to the background expansion of the
universe - i.e. $\alpha$ is no longer constant. In the context of
the Pioneer Anomaly $g_s$ is the gravitational acceleration due to
the Sun.

The time derivative of $\Omega_s$ is then given by equation (15):
\begin{equation}
\dot{\Omega}_s = \frac
{1}{r_s}g_s\frac{v_{rad}}{c}-\frac{1}{r_s}\frac{c^2}{r}\cot{\alpha}
\end{equation}

From equation (4), Hubble's constant ($H$) is identified as the term
($\frac{c}{r}\cot{\alpha}$). Substituting this and $\dot{\Omega}_s$
from equation (15) in equation (9) provides a general result for the
additional acceleration $a_A$ as shown in equation (16). Strictly
speaking, the fifth dimensional parameter, $r$, (i.e. the radius of
curvature of spacetime) is also itself expanding as a result of the
conservation of energy at the boundary of four dimensional
spacetime$^4$, and  in principle this gives rise to a third term in
equation (15). However, the effect is small compared with the other
terms of equation (15). In the context of Earth flybys, the second
term of equation (15) is itself much smaller than the first, which
justifies the use of only the first term in equation (10).
\begin{equation}
a_A = g_s\frac{v_{rad}}{c} - cH
\end{equation}

The Pioneer Anomaly has been seen as the spacecrafts are travelling
away from the Sun, so $g_s$ is negative with respect to the positive
radial velocity, $v_{rad}$ of the spacecrafts - i.e. both terms of
equation (16) have negative values. The same restriction applies to
equation (16) as to equation (10), namely that $a_A$ is not a modification of the gravitational
acceleration due to the Sun. The average radial velocities for
Pioneer 10 at distances from 40\,AU to 70\,AU and for Pioneer 11 at
distances from 22\,AU to 32\,AU were, respectively, 12.6 kms$^{-1}$
and 11.8 kms$^{-1}$ $^5$. Substituting these values in equation (16)
and using a value for $H$ of 71 kms$^{-1}$ Mpc$^{-1}$ and
$M_{\bigodot} = 2\times 10^{30}$\,Kg generates values for $a_A$ as
shown in Table II.

\begin{table}[t]
\title{\bf{Table II: Predicted values for the Pioneer Anomaly
  acceleration.}\label{tab:pion}} \vspace{0.4cm}
\begin{center}
\begin{tabular}{|c|c|}
  \hline
Distance (AU) & $a_A \times 10^{10}$ ms$^{-2}$ \\
\hline \hline Pioneer 11 & \\\hline 22 & 11.7 \\
\hline 27 & 10.1 \\ \hline
32 & 9.2 \\
\hline Pioneer 10 & \\
\hline 40 & 8.5 \\
\hline 50 & 7.9 \\ \hline 60 & 7.6 \\
\hline 70 & 7.4 \\
\hline Average & 9.0 \\
\hline
\end{tabular}
\end{center}
\end{table}

All the predicted figures for $a_A$ lie within the tolerance limits
reported for the Pioneer Anomaly, $8.74\pm 1.33\times
10^{-10}$\,ms$^{-2}$, except for those for Pioneer 11 at distances
less than 27\,AU. However, the most significant feature of equation
(16) is, perhaps, that it predicts a variable value for $a_A$ rather
than the reported constant value.

\section{Discussion}
An analytical framework based on simple assumptions about four
dimensional space-time being closed, isotopic and expanding, and
embedded in a fifth large-scale dimension, $r$, representing the
radius of curvature of space-time, was originally developed by the
authors as a tool to study the orbital dynamics of galaxies [3].
The analysis presented in this paper implies that this framework may
have wider applications. In both cases, the principles of Newtonian
dynamics and gravity are applied without amendment.

For the predicted additional acceleration ($a_A$) to give rise to an
observable effect on a body (i.e. observable within four dimensional
space-time), it must be moving within a system which permits
transfer of kinetic energy to or away from it, such as through
planetary flybys. Moreover, in the case of flybys the net effect of
$a_A$ over the combination of inbound and outbound trajectories will
only be measurable to the extent that these trajectories are
asymmetrical with respect to their energy transfer characteristics.

Several parameters, such as the ecliptic angle of incidence of the
trajectory and its radial velocity towards (or away from) the
planet, determine this asymmetry; however, as Anderson et al.$^1$
have shown, the equatorial latitude appears to be a good surrogate
for this measure of asymmetry, at least in the case of the six Earth
flyby trajectories considered.  The central role of radial velocity
in determining $a_A$ for Earth flybys (equation 10) suggests why no
such velocity discrepancies have been detected for satellites in
bound orbits.

The aggregate force, $F$, due to the combination of gravitational
and the additional acceleration term acting on a mass, $m$, having
radial motion relative to the gravitating body is, using equation
11, given by equation 17, the form of which has an intriguing
similarity to the Lorentz force of electromagnetism.
\begin{equation}
F = m\left(g_s-\frac{c}{r}\wedge \dot{r}\right)
\end{equation}

The equations governing $a_A$ have been derived from first
principles and without the introduction of free parameters or new
constants. That they appear potentially to provide a single explanation for both
Earth flyby and Pioneer anomalies is an added attraction and
justifies their further testing through application in detailed
dynamic models of the solar system.

\section{Addendum}
\subsection{Introduction} 
Although initially promising agreement was achieved between the analytical framework proposed in this paper and the reported flyby anomalies, subsequent detailed modelling undertaken with the assistance of Anderson, suggested the agreement to be much poorer than expected, especially when specific aspects of the Doppler tracking data were considered. Furthermore, the issue of an unidentified energy transfer mechanism (required, if the anomalous acceleration is to be real) remains open, whether energy transfer is correlated to the geocentric latitudes of the spacecraft's inbound and outbound velocity vectors (Anderson et al), so implying that the Earth''s spin energy is the source/sink for energy transferred, or whether the Earth''s orbital energy around the Sun is proposed as the source/sink, as in this paper. 
Evaluation of these results and the issue of energy transfer has led to the conclusion that the  acceleration, $a_A$, derived from consideration of five dimensional dynamics could, alternatively, be interpreted as virtual rather than real. That is, interpreted as an acceleration that affects the orientation of spacecraft's momentum vector but not its energy, so influencing its direction of motion relative to the Earth, rather than its speed. It is the purpose of this addendum to show how this alternative interpretation of $a_A$ may overcome the principal short-comings so far found in the ability of five dimensional dynamics to model accurately the reported flyby anomalies.
\subsection{Interpreting the additional acceleration, $a_A$, as virtual}
In the case of Earth flybys, the additional acceleration, $a_A$, experienced by the spacecraft is directed towards the Earth, in accordance with equation (10), even if it is virtual in nature. However, being virtual it can be treated as analogous to a centripetal acceleration affecting the hyperbolic orbit of the spacecraft about the Earth. Conservation of energy then requires that the effect of $a_A$ on the spacecraft will be such as to induce a change in direction of its motion (effectively a perturbation of the spacecraft''s momentum vector) such that the resultant induced small increase in centrifugal component of acceleration of the spacecraft about the Earth (defined as $a_{cf}$) will be equal and opposite to $a_A$. This principle is illustrated mathematically (and in its simplest form) in equation (18) and diagrammatically in Figure 4. Clearly, however, to the extent that the trajectory of the spacecraft, shown as mass $m$, departs under the influence of $a_A$ from that expected in four dimensional dynamics, its time-dependent profile of kinetic energy and gravitational potential energy inter-change will itself be perturbed and, to that extent, $a_A$ would be expected indirectly to affect the time-dependent kinetic energy profile of the spacecraft as it approaches/recedes from Earth, as a real although second order effect. 

\begin{equation}
a_A + da_{cf}=0
\end{equation} 

\begin{figure}[t]
\begin{center}
\includegraphics[height=3.5in,clip=]{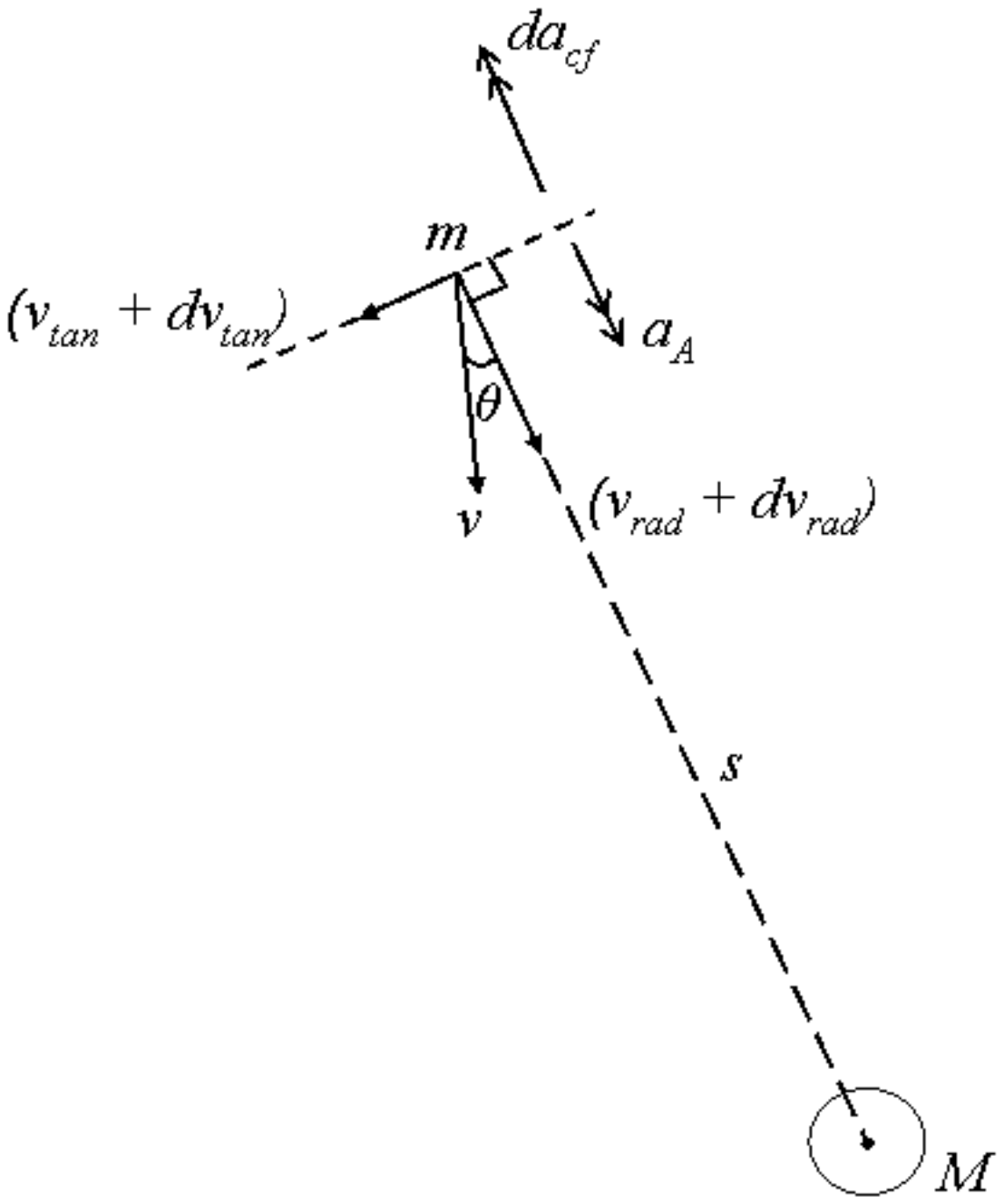}
\vspace*{-50mm}
\caption{Tangential velocity perturbation, $dv_{tan}$, induced by  $a_A$.} \label{fig4}
\end{center}
\end{figure}

Figure 4 considers a point on the trajectory of a spacecraft (mass $m$) in motion towards a flyby of the Earth (mass $M$), and shows the separation of the spacecraft''s velocity, $v$, into two orthogonal components: one, $v_{rad}$,  directed radially towards the Earth and the other, $v_{tan}$, normal to the radial vector, $s$. The centrifugal component of acceleration acting on the spacecraft with respect to the Earth is given by equation (19), where s is the distance between spacecraft and Earth:
\begin{equation}
a_{cf}=\frac{v_{tan}^2}{s}
\end{equation}
 
The proposed effect of $a_A$ is to induce an increase in tangential component of velocity, $dv_{tan}$, of the spacecraft. The gain in tangential component of velocity, $dv_{tan}$, induced by $a_A$ will itself be offset by a reduction in radial component of velocity of the spacecraft with respect to the Earth, $dv_{rad}$, for energy to be conserved under the action of $a_A$, in accordance with equation (20), for $v_{tan} >> dv_{tan}$ and $v_{rad} >> dv_{rad}$. The relationship between an incremental increase in induced centrifugal component of acceleration, $da_{cf}$, and the associated induced incremental increase in tangential component of velocity, $dv_{tan}$, is given by equation (21):
\begin{equation}
v_{tan}dv_{tan}+v_{rad}dv_{rad}=0
\end{equation}
\begin{equation}
\left(a_{cf}+da_{cf}\right)=\frac{\left(v_{tan}+dv_{tan}\right)^2}{s}
\end{equation}
 
From equations (10), (18), (19), (20), (21) and substituting the gravitational acceleration vector, $g_s$, by its component form of $g_s$ = $G$.$M$/$s^2$, for $v_{tan} >> dv_{tan}$ and for $v_{rad} >> dv_{rad} > 0$, it can be shown that the corresponding incremental decrease in the spacecraft radial component of velocity, $dv_{rad}$, is given by the equation:
\begin{equation}
dv_{rad}=-\frac{GM}{2cs}
\end{equation}
 
Care is needed in applying equations (22) because it is valid only for constant spacecraft kinetic energy, as it ignores the time-dependent effect of the Earth's gravitation acceleration on $v$, $v_{rad}$ and $v_{tan}$, in accordance with four dimensional dynamics. Its use is, therefore, strictly limited to being an equation of instantaneous perturbation between four dimensional and five dimensional dynamics. This limitation on equation (22) is especially relevant in the region at or about the flyby itself when, from equation (10), it is apparent that as $v_{rad}$ tends to zero, so does $a_A$ which means that perturbations to the tangential and radial velocity components of the spacecraft, $dv_{tan}$ and $dv_{rad}$ will also tend to zero at the spacecraft perigee. Also, in applying equation (22) it should be noted that we are concerned only with positive values for $dv_{tan}$ and so only with negative values of $dv_{rad}$. Moreover, the relative contributions of $v_{rad}$ and $v_{tan}$ towards the spacecrafts velocity, $v$, vary continuously along the length the spacecraft's inbound and outbound trajectories according to the distance $s$ and angle $\theta$ subtended by the velocity vector $v$ with respect the radial vector. For these reasons numerical integration is required to model accurately the overall effect of $a_A$ on spacecraft trajectory, which is beyond the scope of this paper.  
However, from equation (22) it is possible to see the instantaneous perturbing effect that $a_A$ as a virtual acceleration would have on the spacecraft's trajectory. The spacecraft's inbound trajectory will be slightly skewed away from Earth relative to a trajectory derived by reference solely to four dimensional dynamics (due to a slightly accelerated transition to the purely tangential velocity condition at the perigee), leading to a slightly higher altitude perigee than expected. Conversely, the outbound trajectory will be slightly skewed towards the Earth relative to a trajectory derived by reference solely to four dimensional dynamics. This implies that the perturbing effect of $a_A$ on spacecraft trajectory would, therefore, be asymmetrical about the perigee.    

Lastly, it needs to be stressed that in this analysis the perturbation of tangential velocity, $dv_{tan}$, induced by $a_A$ has been assumed to lie in the direction of the tangential velocity vector itself, $v_{tan}$. This need not necessarily be so, as equation (18) could, in principle, equally well be satisfied by a component of perturbed tangential velocity lying anywhere in the plane normal to the radial vector, s, between spacecraft and Earth. Hence, in the limiting case, equation (18) could even, in principle, be satisfied by an induced gradual gyration of the spacecraft about the radial vector. If this were the case, it may not be directly apparent from the Doppler tracking data which measure spacecraft range and radial velocity only. However, it seems reasonable to proceed on the basis of there being a preferred direction for $dv_{tan}$, not least because the radial motion of spacecraft with respect to the Sun''s gravitational field will itself give rise to a small component of a Sun-directed additional and virtual acceleration, $a_A$, which defines a preferred direction for $dv_{tan}$ relative to the Earth. Finally, it is important to note that equation (18) is not the only possible interpretation of how a virtual acceleration, $a_A$, could induce changes in the direction of the spacecraft velocity vector. It is chosen here for illustration because it is mathematically the simplest.    
\subsection{Discussion}
The above analysis illustrates how $a_A$, although virtual in nature, can give rise to small although systematic and asymmetric perturbations of the inbound and outbound trajectories of a spacecraft during Earth flybys. Since the inference of anomalous changes in spacecraft energy during flybys requires the observed radio Doppler tracking data to be fitted to equations of four dimensional dynamics, it is apparent that such unanticipated perturbations may potentially give rise to apparent anomalies. 
There are potentially two important sets of unknowns in the spacecraft trajectories, as defined by the Doppler tracking data: the first is the tangential component of velocity, $v_{tan}$ which is not directly determined; and the second derives from any absences of reliable Doppler data at or about the perigee. The potential loss of Doppler data during the perigee may derive from a combination of the latitude of spacecraft relative to available tracking stations and difficulties in data quality that can  arise from tropospheric refraction. For example, the gaps in Doppler data for both NEAR(1998)$^1$ and Rosetta(2005)$^6$ flybys are over 1 hour in duration (not necessarily symmetrically centred on the perigee). Both sets of unknowns are deduced from fits of the available Doppler data to the equations of four dimensional dynamics. Since the essence of the proposed additional virtual acceleration, $a_A$, is to induce both a perturbation to the tangential component of spacecraft velocity, $dv_{tan}$, and a slightly higher altitude perigee than expected, there is clearly potential for it to give rise to apparent anomalies in data fitting. 
The task of using Doppler tracking data to infer the extent of trajectory perturbation introduced by $a_A$ may be made harder than expected by the functional form of equation (22). This equation shows that $dv_{rad}$ is proportional to $GM/s$ and so implies that the distance-related  $dv_{rad}$ profile of the spacecraft due to five dimensional effects, should be similar to the distance-related profile of spacecraft gravitational potential energy under four dimensional dynamics, which is also proportional to $GM/s$. The point can be illustrated by considering sections of pre and post encounter trajectories where the approximation $v\approx v_{rad}$ is valid. In these cases, the incremental change in spacecraft velocity ($\Delta$v) at a point along the trajectory can be approximated by the addition of two components: one due to four dimensional dynamics ($\Delta v_{4d}$); and the other due to five dimensional effects ($\Delta v_{5d}$)\footnote{$\Delta v_{5d}\approx  dv_{rad}\left(s-\Delta s\right)-dv_{rad}\left(s\right)$ from equation (22)} – the latter being taken from equation (22). This approximation is shown in equation (23) and is rearranged in equation (24)\footnote{In equation (24), the sign will be reversed for outbound trajectories}, for $s >> \Delta s$, to highlight the way in which five dimensional effects would show-up as a perturbation ($\epsilon_p$) to the four dimensional case. 
\begin{equation}
\Delta v \approx \Delta v_{4d}+\Delta v_{5d} \approx \Delta v_{4d}\left(1+\epsilon_p\right)
\end{equation}                                                           \begin{equation}
\Delta v \approx \frac{GM\Delta s}{vs^2}\left(1-\frac{v}{2c}\right)
\end{equation}                                                                        
In fitting the Doppler data to four dimensional equations of motion, a time dependent perturbation introduced by the term $v/2c$ may be absorbed by the imputed changing values of $\Delta v$ (and so $v$) as an underestimate of the implied value of the tangential component of velocity of the spacecraft, which is not independently measurable. Depending on the step length, $\Delta s$, used in the numerical integration, the perturbation could fall within the error bars of the methodology where the approximation $v \approx v_{rad}$ is valid. For example, applying equation (24) for a spacecraft at a distance from Earth, $s$, of $10^5$ km, an integration step length, $\Delta s$, of $\sim10^3$ km or less (equivalent to 100 seconds or less, at a velocity of 10 kms$^{-1}$) would be consistent with a fit to four dimensional dynamics within an equivalent  noise level of less than $\sim$0.1\,mms$^{-1}$ in spacecraft velocity. Consequently, both inbound and outbound limbs of the spacecraft's trajectory could provide good fits to four dimensional dynamics and yet be {\em inconsistent} with each other. This inconsistency could show up, inter alia, as an un-modelled change in spacecraft's energy. Such an effect would be consistent with the anomaly becoming immediately apparent, within the fitted trajectories, from the point at which the spacecraft emerges from its encounter with the Earth and radio tracking contact is resumed\footnote{A similar effect will be expected in fitting trajectories to 4-dimensional dynamics if  $a_A$ were interpreted as real and $\Delta v_{5d}$ were taken from equation (13) rather than (22)}.  
 
Moreover, the rate of decline of $a_A$ with spacecraft distance from Earth evident from equation (10), and the inverse dependency of $\Delta v_{rad}$ on spacecraft distance from Earth evident in equation (22) would imply a strong distance dependency for difficulties in achieving consistent fits to four dimensional dynamics, in cases where the proposed five dimensional perturbation effects arise. The consequent reducing potential for reported anomalies in cases of higher altitude flybys is consistent with the conclusions of Anderson and Nieto [7].  

In particular, equations (23) and (24) can be rearranged to estimate the altitude, $H_{limit}$, below which the perturbation $\Delta v_{5d}$ will exceed a specified sensitivity level (e.g. 0.1\,mms$^{-1}$) as follows:
\begin{equation}
H_{limit}\approx \left(\frac{GM}{2c}\frac{|\Delta s|}{|\Delta v_{5d}|}\right)^{1/2}-R_{Earth}
\end{equation} 
For a spacecraft velocity $\sim10\,$kms$^{-1}$ and integration step length of 1\,s, $H_{limit}$ is then $\sim 2000\,$km. For a given satellite  $H_{limit}$ will depend on its detailed trajectory.  However equation (25) does indicate the altitude above which anomalies would not be detectable, and is in general agreement with observation to date [7].       
 
It can be further noted that whereas the interpretation of $a_A$ as real, and acting in a radial direction towards Earth, may present difficulties in interpreting anomalous changes in spacecraft angular momentum during the flyby, the interpretation of $a_A$  as virtual may not, as it concerns the induction of perturbations to the spacecraft velocity vector but not its energy. 

The interpretation of $a_A$ as virtual is not specific to Earth flybys, so some comment can also be made about the Pioneer Anomaly and equation (16). This equation has two components, the first ($g_s$$v_{rad}$$/c$) derived from five dimensional dynamics, and the second ($-cH$) derived from consideration of the background expansion of a five dimensional universe. Whereas the first component has a clear virtual analogue inducing changes in spacecraft momentum vector but not in its energy (so causing the instantaneous tangential component of Pioneer spacecraft velocity to increase at the expense of the radial component), the second component of equation (16)\ does not.  Considering only the first term, therefore, and for a spacecraft on a hyperbolic orbit travelling away from the Sun, it is possible to use equation (22) to derive a time derivative of the perturbation to the radial velocity, $dv_{rad}$, of the spacecraft that is induced by the virtual acceleration, $a_A$, relative to that expected under four-dimensional dynamics. The acceleration, $a_A$, is in this case directed towards the Sun. The time derivative is given by equation (26) which is simply half the effect of that implied if $a_A$ were treated as real (i.e. half the first term of equation (16)). 
\begin{equation}
\frac{dv_{rad}}{dt}=-\frac{g_sv_{rad}}{2c}
\end{equation} 
The values of the Sun-directed acceleration computed from equation (26) for Pioneers 10 and 11, lie in the range 0.3 to $2.4\times 10^{-10}\,$ms$^{-2}$ which is well below the reported values implying that this effect alone cannot account for the anomaly [8,9].
\subsection{Conclusions}
This Addendum has been included to address the possibility of the additional acceleration, $a_A$, derived from five dimensional dynamics, being interpreted as {\em virtual} rather than real. The analytical framework used is otherwise unchanged from that in the body of this paper. The next step, as before, is to carry out numerical integration of spacecraft equations of motion modified for the perturbing effects of $a_A$ and to reconcile these results with trajectory fits to available Doppler tracking data, to see whether these perturbations are consistent with the reported flyby and Pioneer anomalies.

\end{document}